\documentclass[journal]{IEEEtran}
\usepackage{bm}
\usepackage{amsfonts}
\usepackage{amsmath}
\usepackage{graphics}

\begin{document}
\title{Transmission Capacity of Ad Hoc Networks\\ with Spatial Diversity
\thanks{This work was supported by National Instruments and by the
NSF under grant no. 0635003 (Weber), no. 0634979 (Andrews), and the
DARPA IT-MANET program, Grant W911NF-07-1-0028. A portion of this
work was presented at ISIT '07 in Nice, France.
\newline \indent A. M. Hunter and J. G. Andrews are with the Wireless
Networking and Communications Group (WNCG) of the Electrical and
Computer Engineering Department, The University of Texas at Austin,
Austin, TX, 78712-0240 USA (email: \{hunter,
jandrews\}@ece.utexas.edu). \newline \indent S. Weber is with the
Department of Electrical and Computer Engineering, Drexel
University, Philadelphia, PA 19104-2875 USA (email:
sweber@ece.drexel.edu).}}
\author{Andrew M. Hunter, \IEEEmembership{Student Member, IEEE},
Jeffrey G. Andrews, \IEEEmembership{Senior Member, IEEE}, and Steven
Weber, \IEEEmembership{Member, IEEE}} \maketitle

\begin{abstract}
This paper derives the outage probability and transmission capacity
of ad hoc wireless networks with nodes employing multiple antenna
diversity techniques, for a general class of signal distributions.
This analysis allows system performance to be quantified for fading
or non-fading environments. The transmission capacity is given for
interference-limited uniformly random networks on the entire plane
with path loss exponent $\alpha>2$ in which nodes use: (1) static
beamforming through $M$ sectorized antennas, for which the increase
in transmission capacity is shown to be $\Theta(M^2)$ if the
antennas are without sidelobes, but less in the event of a nonzero
sidelobe level; (2) dynamic eigen-beamforming (maximal ratio
transmission/combining), in which the increase is shown to be
$\Theta(M^{\frac{2}{\alpha}})$; (3) various transmit antenna
selection and receive antenna selection combining schemes, which
give appreciable but rapidly diminishing gains; and (4) orthogonal
space-time block coding, for which there is only a small gain due to
channel hardening, equivalent to Nakagami-$m$ fading for increasing
$m$. It is concluded that in ad hoc networks, static and dynamic
beamforming perform best, selection combining performs well but with
rapidly diminishing returns with added antennas, and that space-time
block coding offers only marginal gains.
\end{abstract}


\section{\label{intro}Introduction}

\PARstart{P}{rior} work on ad hoc network capacity has focused on
the limiting behavior as the network grows large.  For the purpose
of ascertaining the effect that multiple antennas has on the
capacity of the network, the more pertinent question is how the
capacity scales with the number of antennas at each node. Naturally,
this scaling will differ depending on the way that the antennas are
utilized. The goal of this paper is to determine which
multiple-antenna techniques perform best in a given network density
or similarly, which technique can support the densest network.

Multi-antenna systems (MIMO) are currently of great interest in all
wireless communication systems due to their potential to combat
fading, increase spectral efficiency, and potentially reduce
interference. Over the past decade, many different MIMO techniques
have been proposed, which can be grouped into three broad
categories: diversity-achieving, beam-steering, and spatial
multiplexing. Diversity-achieving techniques increase reliability by
combatting or exploiting channel variations. Beam-steering
techniques increase received signal quality by focusing desired
energy or attenuating undesired interference. Spatial multiplexing
aggressively increases the data rate by transmitting independent
data symbols across the antenna array. In this paper we focus on the
first two types of techniques, which do not increase the number of
independent datastreams and hence are easier to fairly compare.  We
also expect that these techniques will be more relevant than spatial
multiplexing in interference-limited ad hoc networks since sending a
single datastream in low-SNR links is superior in terms of both
performance and implementation complexity \cite{IEEEbib:raohassi},
\cite{IEEEbib:choiand}. This paper develops a framework for
comparing the utility of the diversity-providing and beam-steering
MIMO techniques, with the goal of providing insight on how to use
multiple antennas in ad hoc networks.

\subsection{Background and Related Work}

Recent advances in characterizing network capacity were sparked by
\cite{IEEEbib:gk} with its notion of transport capacity and a number
of works have followed in the same vein including
\cite{IEEEbib:xieprk}, \cite{IEEEbib:levtel}, and \cite{ozlevtse}.
These studies focus on the behavior of end-to-end network capacity
in the limit as the number of nodes grows large under a variety of
models of node interaction and fading conditions. These confirm the
basic intuition from \cite{IEEEbib:gk} that, under traditional
technological or physical limitations on node cooperation and signal
reception, transmissions require ``area'' in which to take place and
so per node end-to-end throughput decays as
$\Theta(\frac{1}{\sqrt{n}})$ for $n$ nodes in the network. A
fundamental change occurs with significant mobility as
\cite{IEEEbib:grodste} and \cite{IEEEbib:neelmod} show since optimal
routing can take on new forms to tradeoff throughput and delay.

An alternative characterization of ad hoc network capacity was
developed in \cite{IEEEbib:ag2} which defined rate regions for given
network configurations and traffic needs. This was extended to the
MIMO case in \cite{IEEEbib:sb} and the notion of ``capacity region''
was extended in several ways. This versatile approach has the
drawback of being prohibitively computationally intensive for
analyzing large networks. It also focuses on large network
optimization problems that would be difficult to solve in a
distributed system at present.

A straightforward way to evaluate a physical layer technique under
per node service requirements is to determine the maximum density of
concurrent transmissions, or the \emph{optimal contention density},
for which each node's requirements are still met. This leads
naturally to the transmission capacity metric which is defined in
\cite{IEEEbib:tc} to be the maximum allowable spatial density of
successful transmissions multiplied by their data rate given an
outage constraint. For an outage constraint $\epsilon$ and a
transmission data rate $b$ in bits/s/Hz or per channel use, the
transmission capacity is given by
$c_\epsilon=b(1-\epsilon)\lambda_\epsilon$ for the optimal
contention density $\lambda_\epsilon$. The transmission capacity is
then the area spectral efficiency resulting from the optimal
contention density.

Computing the transmission capacity is made possible by using a
spatial point process to model node positions, as pioneered in the
analysis of wireless networks by \cite{IEEEbib:klesilv}. More
recently, Haenggi, \emph{et al.} in \cite{IEEEbib:Haenggi},
\cite{IEEEbib:HaenggiLiu}, and \cite{IEEEbib:HaenggiGanti}
emphasized the importance of network topology by characterizing some
of the distinctions in throughput, interference, and outage in
regular as well as clustered random networks. This approach was also
taken in \cite{IEEEbib:WAJeffect} which developed bounds on the
transmission capacity for general fading models as well as power
control and scheduling schemes with only individual channel state
information and single antennas.

Several papers, including \cite{IEEEbib:sh}, \cite{IEEEbib:ka},
\cite{IEEEbib:ka2}, have studied the effects of cochannel
interference on MIMO. However, these studies lack a clear link
between point-to-point throughput and network performance gains. It
is presently unclear which MIMO technologies yield the highest gains
in large random networks. For example, \cite{IEEEbib:bl} uses a
game-theoretic analysis to show that capacity is maximized for
mutually interfering sources when each sends only one datastream,
while \cite{IEEEbib:cg} and \cite{IEEEbib:jj} suggest capacity is
improved through spatially multiplexing potentially multiple
transmissions; however, \cite{IEEEbib:cg} again focuses on
asymptotics in the number of nodes and the results of
\cite{IEEEbib:jj} are obscured by the mobility/delay issue.

\subsection{Contributions}

This paper analyzes networks with single-datastream MIMO diversity
techniques including beamforming, antenna sectorization, space-time
block coding, and selection combining in Rayleigh fading, terms
which we will make precise in the course of the paper. The gains in
transmission capacity of each are shown and compared, especially as
a function of the number of antennas. Results are also given for
Nakagami-$m$ fading for integer $m$ to compare methods in
line-of-sight versus non-line-of-sight propagation environments and
to assist in interpreting channel hardening gains. While spatial
multiplexing techniques are omitted, they are left as future work
though some of the results developed here will also be applicable to
spatial multiplexing systems. Since diversity techniques are robust
in noise-limited environments and generally reasonable to implement,
they constitute an important subset of the primary MIMO techniques.
Also, as indicated in \cite{IEEEbib:bl} a game-theoretic analysis
indicates optimality of single stream techniques in interference
limited environments.


The goal of the paper will be to establish several clear relations
between the optimal contention density and the number of antennas,
which we list below. More precisely, for random wireless networks on
the entire plane using the above MIMO techniques in block fading
channels with path loss, this paper determines transmission
capacity, the scaling of the optimal contention density with the
number of antennas, and outage probabilities as a function of
network parameters. Under small outage constraints, for $M_t$ and
$M_r$ the number of transmit and receive antennas, respectively, and
$\alpha>2$ is the path loss exponent, we have:
\begin{enumerate}
\item Ideal Sectorized Antennas: $\lambda_\epsilon = \Theta(M_tM_r)$
\item Sectorized Antennas with sidelobe level $\gamma\in [0,1]$:
$\lambda_\epsilon =
\Theta\left[\left(\frac{M}{1+\gamma^{\frac{2}{\alpha}}(M-1)}\right)^2\right]$
for $M=M_t=M_r$
\item Maximal Ratio Combining (MRC): $\lambda_\epsilon = \Theta(M_r^{\frac{2}{\alpha}})$
\item Maximal Ratio Transmission (MRT) and Combining:
$\lambda_\epsilon=O((M_tM_r)^{\frac{2}{\alpha}})$,
$\lambda_\epsilon=\Omega(\max\{M_t,M_r\}^{\frac{2}{\alpha}})$
\item Orthogonal Space-Time Block Coding (OSTBC): $\lambda_\epsilon =
\Theta(M_r^{\frac{2}{\alpha}})$.
\end{enumerate}

The orderwise results demonstrate that as spatial diversity
techniques increase the SINR, network throughput increases better
than logarithmically in the number of antennas. In particular, these
relations demonstrate that beamforming, either static or dynamic,
achieves the most network transmission capacity increase among
diversity techniques. On the other hand, space-time block coding
yields little, especially for more antennas than two. The results
also highlight the advantages of achieving diversity at the receiver
since open loop transmit diversity techniques have some specific
drawbacks to be discussed and receiver techniques do not require
feedback.\newline

The remainder of the paper is organized as follows: Section
\ref{sec:model} presents the network model and derives properties
for Poisson shot noise functionals applicable to large class of MIMO
techniques. Section \ref{sec:nakag} discusses the optimal contention
density for single antenna systems in Nakagami-$m$ fading. The
optimal contention density for networks of nodes with multiple
sectorized antennas is derived in Section \ref{sec:sector}. Section
\ref{sec:MRC} derives the outage probabilities and optimal
contention densities for receive MRC systems and Section
\ref{sec:MRT} does the same for MIMO MRT/MRC systems. Networks using
OSTBCs are analyzed in Section \ref{sec:ostbc}. Section
\ref{sec:selcomb} gives transmission capacity results for selection
combining in ad hoc networks and Section \ref{sec:concl} concludes.

\section{\label{sec:model}The Network Model and Analytical Methods}

\subsection{The Model}

This section defines the network model and presents some results on
Laplace functionals of Poisson shot noise processes which will be
used in later sections. In order to focus on the physical layer,
consider a wireless network operating a random access protocol in
the style of slotted ALOHA without power control. As discussed in
\cite{IEEEbib:tc}, this model includes the collision behavior of
practical distributed systems while neither addressing nor
precluding the issue of routing. It also provides a way to give a
clear relationship between the network throughput and the number of
antennas employed for each technique. Let the distribution of
transmitting nodes in the network be a stationary marked Poisson
point process with intensity $\lambda$ in $\mathbb{R}^2$; the
process is denoted by $\Phi$. To analyze the performance of a random
access wireless network, consider a typical receiver located at the
origin. As a result of Palm probabilities of a Poisson process,
conditioning on the event of a node lying at the origin does not
affect the statistics of the rest of the process (see
\cite{IEEEbib:dg}, ch. 2). Moreover, due to stationarity of the
Poisson process, the statistics of signal reception at this receiver
are seen by any receiver.

To model propagation through the wireless channel, let signals be
subject to path loss attenuation model $d^{-\alpha}$ for a distance
$d$ with exponent $\alpha>2$ as well as small scale fading for
either a Rayleigh or Nakagami-$m$ fading distribution with unit
mean. Also, let all nodes transmit with the same power $\rho$. For
such a channel, the typical receiver obtains desired signal power
$\rho S_0R^{-\alpha}$ for some fixed transmitter-receiver separation
distance $R$, and with a fading power factor $S_0$ on the signal
from its intended transmitter, labeled $0$. The interfering nodes,
numbered $1, 2, 3,...$ constitute the marked process
$\Phi=\{(X_i,S_i)\}$, with $X_i$ denoting the location of the $i$th
transmitting node, and with marks $S_i$ that denote fading factors
on the power transmitted from the $i$th node and then received by
the typical receiver. Thus the receiver receives interference power
$\rho S_i|X_i|^{-\alpha}$ from the $i$th interfering node ($|\cdot|$
denoting magnitude). For single-antenna narrowband systems in
Rayleigh fading channels, for example, the power factors $S_0$ and
$S_i$ are distributed exponentially with unit mean so that the mean
interfering power is governed by transmit power and path loss.

Successful transmission occurs if the inequality
\begin{equation}
\label{eq:suc}
\frac{\rho S_0 R^{-\alpha}}{\rho I_\Phi+N_0}\geq \beta
\end{equation}
is satisfied for some target signal-to-interference-and-noise ratio
(SINR) $\beta$, aggregate co-channel interference $\rho I_\Phi$, and
thermal noise $N_0$. The aggregate interference is a Poisson shot
noise process (scaled by $\rho$), which is a sum over the marked
point process:
\begin{equation}
\label{eq:inter} I_\Phi=\sum_{X_i\in\Phi}S_i|X_i|^{-\alpha}
\end{equation}
with $|X_i|$ denoting the distance of $X_i$ from the origin. From
here on, it will be assumed that the network is interference
limited, with $\rho I_\Phi \gg N_0$ so that thermal noise is
negligible. Following \cite{IEEEbib:ba}, the probability of
successful transmission for a typical receiver is:
\begin{eqnarray}
\mathbf{P}(SIR\geq\beta)&=&\mathbf{P}\left(\frac{\rho S_0
R^{-\alpha}}{\rho I_\Phi}\geq \beta\right)=\mathbf{P}\left(S_0\geq
\beta R^{\alpha}
I_\Phi\right)\nonumber\\
&=&\int_{0}^{\infty} {\mathbf P}(S_0\geq s\beta
R^{\alpha})\;f_{I_\Phi}(s)\mathrm{d}s \nonumber\\
&=&\int_0^\infty F^c_{S_0}(s\beta R^{\alpha})
f_{I_\Phi}(s)\mathrm{d}s \label{eq:outagedef}
\end{eqnarray}
where the third step is reached by conditioning on $s$ and
$F^c(\cdot)$ denotes a complementary cumulative distribution
function (CCDF). In the single antenna (SISO) case, the received
signal power is exponentially distributed with $F^c_{S_0}(s\beta
R^{\alpha})=e^{-s\beta R^{\alpha}}$ so that
\begin{equation}
\label{eq:lap} \mathbf{P}(SIR\geq\beta)=\int_0^{\infty} e^{-s \beta
R^{\alpha}}\; f_{I_\Phi}(s)\mathrm{d}s \; .
\end{equation}
This is now a Laplace transform of the PDF of $I_{\Phi} $ which
gives $\mathbf{P}(SIR\geq\beta) = \mathcal{L}_{I_\Phi }(\beta
R^\alpha)$. The Laplace transform for a general Poisson shot noise
process in $\mathbb{R}^2$ with independent, identically distributed
(i.i.d.) marks $S_i$ is given by \cite{IEEEbib:kp}
\begin{equation}
\label{eq:lapl} \mathcal{L}_{I_\Phi }(\zeta)=\mathrm{exp}
\left\{-\lambda\int_{\mathbb{R}^2}1-E\left [e^{-\zeta S
|x|^{-\alpha}}\right ]\mathrm{d}x\right\}
\end{equation}
where the expectation, denoted by $E[\cdot]$, is over $S$ which has
the same distribution as any $S_i$. Note that we are using the
simplified attenuation function $|d|^{-\alpha}$. While this model is
inaccurate in the near field, most notably because it explodes at
the origin, for systems operating primarily in the far field (e.g.,
$R$ is many carrier wavelengths), this inaccuracy has negligible
effect for the purpose of calculating outage probabilities. One can
modify the path loss function to $\frac{1}{1+|d|^\alpha}$, for
example, as mentioned in \cite{IEEEbib:ba} and perform the same
analysis. The result is that for $R$ well in the far field, this
modification leads to the same transmission capacity conclusions
though with more cumbersome analysis.

For Rayleigh fading channels, i.e., $S_0,S_i\sim \mathrm{Exp}(1)$,
(\ref{eq:lapl}) simplifies to
\begin{equation}
\mathcal{L}_{I_\Phi}(\zeta)=\mathrm{exp}
\left\{-2\pi\lambda\int_{0}^{\infty}\frac{u}{1+
|u|^{\alpha}/\zeta}\mathrm{d}u\right\}=e^{-\lambda C
\zeta^{\frac{2}{\alpha}}}
\end{equation}
with $\mathcal{L}_{I_\Phi}(\zeta)$ evaluated at $\zeta=\beta
R^\alpha$ and
$C=\frac{2\pi}{\alpha}\Gamma(\frac{2}{\alpha})\Gamma(1-\frac{2}{\alpha})$,
with $\Gamma(t)=\int_0^\infty x^{t-1}e^{-x}\mathrm{d}x$ being the
Gamma function. Note that in general $C$ depends on $S_i$, and so
for some systems, it may no longer be a function of the path loss
exponent alone. For all cases considered in this paper, the integral
in (\ref{eq:lapl})
\begin{equation*}
\int_{\mathbb{R}^2}1-E\left [e^{-\zeta S_i |x|^{-\alpha}}\right
]\mathrm{d}x
\end{equation*}
where $\zeta$ is evaluated at $\beta R^\alpha$, will be proportional
to $(\beta R^\alpha)^{\frac{2}{\alpha}}$. This has the simple sphere
packing interpretation that each transmission takes up an ``area''
proportional to $(\beta^{1/\alpha}R)^2$.

\subsection{The Optimal Contention Density}

Applying a small outage constraint (e.g., $\epsilon<.1$) to
(\ref{eq:lap}), the network just meets this constraint when
\begin{equation}
\mathbf{P}(SIR\geq\beta)=1-\epsilon=e^{-\lambda C
R^2\beta^{\frac{2}{\alpha}}}
\end{equation}
and solving for $\lambda$ yields the optimal contention density:
\begin{equation}
\label{eq:linearize} \lambda_\epsilon=\frac{-\ln(1-\epsilon)}{C
R^2\beta^{\frac{2}{\alpha}}} =\frac{\epsilon}{C
R^2\beta^{\frac{2}{\alpha}}}+\Theta(\epsilon^2) \; .
\end{equation}
Since the results herein will focus on the small outage regime, it
will be convenient to introduce the notation
$\lambda_\epsilon=\bar{\lambda_\epsilon}+O(\epsilon^2)$ allowing
equations to be expressed in terms of $\bar{\lambda_\epsilon}$ with
the $O(\epsilon^2)$ error terms merely implied. The result
(\ref{eq:linearize}) given in \cite{IEEEbib:tc} and
\cite{IEEEbib:ba} can be generalized through the following Theorem.

\paragraph*{Theorem 1} \emph{Let the interfering transmitters form a Poisson
process of intensity $\lambda$ around a typical receiver with the
outage probability being $\mathbf{P}(SIR\leq
\beta)=\mathbf{P}(\frac{\rho S_0 R^{-\alpha}}{\rho
I_\Phi}\leq\beta)$ with fixed $\rho$, $\beta$, $R$, and $\alpha$.
Suppose $F^c_{S_0}$ takes the form
\begin{equation}
F^c_{S_0}(x)=\sum_{n\in\mathcal{N}} e^{-n x}\sum_{k\in\mathcal{K}}
a_{nk} x^k
\end{equation}
for finite sets\footnote[1]{Note that not all sets lead to valid
distributions, e.g., for $n=k=1$, $F^c(x)\propto xe^{-x}$ which
cannot be a valid CCDF. Hence, the expressions given in the Theorem
rely on a valid CCDF to be correct.}
$\mathcal{N,K}\subset\mathbb{N}$, and suppose $S_0$ is independent
of $I_\Phi$, then
\begin{equation}
\label{eq:PSIR} \mathbf{P}(SIR\geq\beta) = \sum_{n\in\mathcal{N}}
\sum_{k\in\mathcal{K}} \left[a_{nk} \left(-\frac{\zeta}{n}\right)^k
\frac{d^k}{d\zeta^k}
\mathcal{L}_{I_\Phi}(\zeta)\right]_{\zeta=n\beta R^\alpha}.
\end{equation}
Furthermore, for a small outage constraint $\epsilon$, the optimal
contention density is given by:
\begin{equation}
\bar{\lambda_\epsilon}=\frac{K_\alpha}{C_\alpha}\frac{\epsilon}{R^2
\beta^{\frac{2}{\alpha}}}
\end{equation}
for
\begin{equation}
\label{eq:K} K_\alpha=\left[\sum_{n\in\mathcal{N}}
\sum_{k\in\mathcal{K}}
a_{nk}n^{\frac{2}{\alpha}-k}\prod_{l=0}^{k-1}(l-2/\alpha)\right]^{-1}
\end{equation}
and $C_\alpha(\beta
R^{\alpha})^{\frac{2}{\alpha}}=\int_{\mathbb{R}^2}1-E[e^{-\zeta S_i
|x|^{-\alpha}}]\mathrm{d}x$.}

\begin{proof}The proof is presented in Appendix \ref{sec:pflem1}.\end{proof}

This Theorem has two key contributions and several implications. The
Theorem's two contributions are (1) that it gives the exact
probability of outage for any network density or target SIR (thermal
noise was eliminated for convenience but could easily be reinserted,
see Appendix \ref{sec:pflem1}) but also (2) that it gives a solution
for the optimal contention density in the low outage regime. As for
consequences of the Theorem, first, it reinforces the linear
dependence of the optimal contention density with the outage
constraint for uniformly distributed random access systems. When
compared with a regular network topology, $\epsilon$ essentially
becomes a penalty factor on the area spectral efficiency achievable
with random access. Second, it shows that a large class of received
signal and fading distributions is amenable to a transmission
capacity analysis, including a number of MIMO techniques. Third, it
demonstrates that derivation of the transmission capacity consists
of two components: (1) determining $K_\alpha$ which is dependent on
the received signal distribution, and (2) determining $C_\alpha$
which is a result of the interfering signal statistics. This holds
in general only when the condition of independence between the
received signal distribution and the interfering shot noise process
is satisfied.

The results in Theorem 1 give fundamental limits on the operating
point of a communicating pair and its performance in an
interference-limited environment and there are several ways one
could interpret these expressions. One interpretation is that for a
communicating pair amidst a density $\bar{\lambda}$ of interferers,
the pair is free to choose any rate-outage-distance operating point
for which $\bar{\lambda}\leq\frac{K_\alpha\epsilon}{C_\alpha
\beta^\frac{2}{\alpha}R^2}$. Furthermore, the operating point can be
chosen independently of the operating points of any other pair and
hence the statement of the Theorem is very general.

On the other hand, if network-wide performance constraints $\beta$
and $\epsilon$ are imposed, then implicitly an upper limit on $R$ is
also established. If amidst a given density $\bar{\lambda}$ a pair
of nodes wish to communicate over a distance greater than $R$, then
either outage probability or data rate must suffer. For clarity of
the presentation, we will assume that $\epsilon$, $\beta$, and $R$
are fixed for every communicating pair. The analysis could be
expanded by permitting a distribution on $R$ in which case:
\begin{equation*}
\bar{\lambda}=\frac{K_\alpha\epsilon}{C_\alpha
\beta^\frac{2}{\alpha}}\int\frac{1}{r^2}\mathrm{d}F_R(r)
\end{equation*}
for which the maximum distance still permits the small outage
approximation. It was shown in \cite{IEEEbib:WAJeffect} that
considering variable transmission distances has minimal impact on
the transmission capacity. Specifically, the transmission capacity
is reduced by the factor $E[R^2]/E[R]^2$ when one imposes a
distribution on $R$. Finally, the benefit in terms of transmission
capacity to the network of the various MIMO techniques (embodied in
the factors $K_\alpha$ and $C_\alpha$) remain unaffected by variable
transmission distances.

\section{\label{sec:nakag}Transmission Capacity in LOS and NLOS Environments}
In \cite{IEEEbib:tc}, the same Poisson network model was used but
propagation was modeled with path loss only while \cite{IEEEbib:ba}
incorporated Rayleigh fading in addition to path loss. In order to
characterize the effect on network capacity between these extremes,
Rayleigh fading and non-fading, let the envelope of the received
signal be Nakagami-$m$ distributed with integer parameter $m$ in
addition to being scaled by path loss. The Nakagami distribution
includes Rayleigh as a special case ($m=1$), non-fading as a special
case ($m=\infty$), and provides a close parameterized fit for
empirical data as well as the Ricean distribution for
$m=\frac{(K+1)^2}{(2K+1)}$ for $K$ the Ricean factor
\cite{IEEEbib:gs}. Theorem 1 is applied as follows:
\paragraph*{Proposition 1} \emph{For a random access single-antenna narrowband
wireless network in Nakagami-$m$ fading for $m\in \mathbb{N}$, the
optimal contention density with outage $\epsilon$ is given by}
\begin{equation}
\label{eq:naklamthm}
\bar{\lambda_\epsilon}=\frac{K_{\alpha,m}\epsilon}{C_{\alpha,m}\beta^{\frac{2}{\alpha}}R^2}
\end{equation}
\emph{where}
\begin{equation}
\label{eq:kmrcP}
K_{\alpha,m} =
\left[1+\sum_{k=0}^{m-2}\frac{1}{(k+1)!}
\prod_{l=0}^{k}(l-2/\alpha)\right]^{-1}
\end{equation}
\emph{and}
\begin{equation}
\label{eq:CP}
C_{\alpha,m}=\frac{2\pi}{\alpha}\sum^{m-1}_{k=0}{m\choose k}
B\left(\frac{2}{\alpha}+k;m-\left(\frac{2}{\alpha}+k\right)\right)
\end{equation}
\emph{with $B(a,b)=\frac{\Gamma(a)\Gamma(b)}{\Gamma(a+b)}$ being the
Beta function. Further, both $K_{\alpha,m}$ and $C_{\alpha,m}$
increase as $\Theta(m^{\frac{2}{\alpha}})$ with
\begin{equation}
\label{eq:Kbound} m^{\frac{2}{\alpha}}\leq K_{\alpha,m}\leq
\Gamma(1-2/\alpha)m^{\frac{2}{\alpha}}
\end{equation}
and}
\begin{equation}
0<\frac{K_{\alpha,1}}{C_{\alpha,1}}<\frac{K_{\alpha,m}}{C_{\alpha,m}}<\frac{1}{\pi}.
\end{equation}

\begin{proof}The proof is presented in Appendix
\ref{sec:pfThm1}.\end{proof}

\begin{figure}
\begin{center}
\resizebox{9cm}{7cm}{\includegraphics{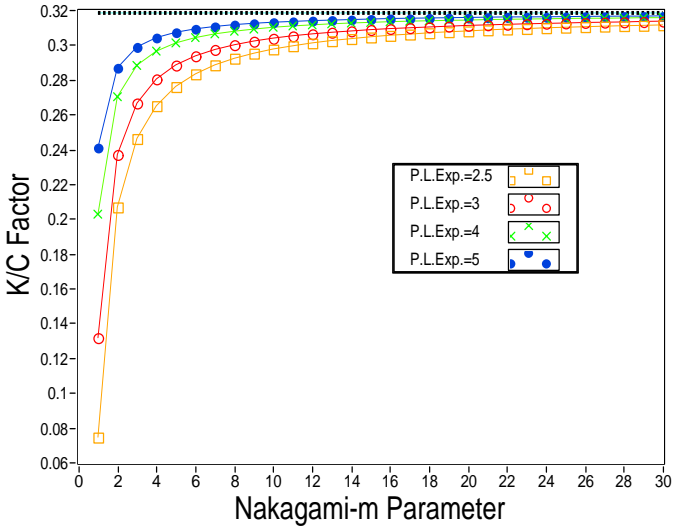}}
\caption{$\frac{K_{\alpha,m}}{C_{\alpha,m}}\rightarrow .318
\approx\frac{1}{\pi}$ as $m\rightarrow\infty$ for various path loss
exponents.} \label{fig:nakKC}
\end{center}
\end{figure}

In \cite{IEEEbib:tc} the transmission capacity in a non-fading
environment is bounded above by
$\bar{\lambda_\epsilon}\leq\frac{\epsilon}{\pi\beta^{\frac{2}{\alpha}}R^2}$
and further more this upper bound is fairly tight which implies that
in fact $\lim_{m\rightarrow
\infty}\frac{K_{\alpha,m}}{C_{\alpha,m}}\approx\frac{1}{\pi}$. Fig.
\ref{fig:nakKC} shows the ratio $\frac{K_{\alpha,m}}{C_{\alpha,m}}$
for various $\alpha$ versus $m$. This reinforces the (rough)
tightness of the upper bound in \cite{IEEEbib:tc}.

Proposition 1 bridges the gap between fading and non-fading
environments and demonstrates the potentially significant gain in
network capacity relative to non-fading environments. It also shows
that environments with lower path loss suffer more from severe
fading (when in the common practical case $\beta>1$) and improve
more with a strong LOS. The distinction is particularly important
for dense networks communicating with nearby neighbors which are
likely to have lower path loss \emph{and} a significant LOS. The
results also reveal the gains to be reaped by diversity techniques
that can mitigate fading. The particular results for $K_{\alpha,m}$
and $C_{\alpha,m}$ will also be significant when analyzing MIMO
techniques.

\section{\label{sec:sector}Sectorized Antennas}

\begin{figure}
\begin{center}
\resizebox{7.5cm}{6cm}{\includegraphics{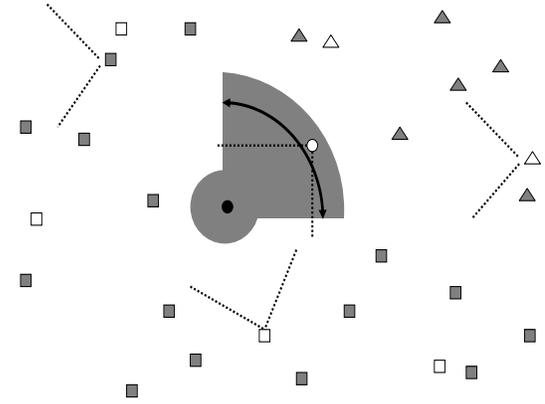}}
\caption{Sectorized antenna model with a $\frac{\pi}{2}$ sector main
beam and constant sidelobe level for the receiver of interest (black
dot), its intended transmitter (white dot), and the four sets of
interferers: $\Phi_1$ the interferers in the active sector
transmitting toward the receiver (white triangles), $\Phi_2$ the
interferers in the active sector transmitting away from the receiver
(shaded triangles), $\Phi_3$ the interferers out of the active
sector transmitting toward the receiver (white squares), and
$\Phi_4$ the interferers out of the active sector transmitting away
from the receiver (shaded squares).} \label{fig:sectmodel}
\end{center}
\end{figure}

Now consider the same network model but with transmitters and
receivers that are each equipped with $M$ sectorized antennas. Let
each antenna cover an angle of $\frac{2\pi}{M}$ radians with an
aperture gain of $M$ for both transmitting and receiving in its
sector and (potentially) with some small input/output gain outside
its sector. Assume each transmitter picks a receiver in a uniformly
random direction, and for each transmitter/receiver pair both know
the sector in which to communicate with their intended partner. The
model can include a constant sidelobe level $\gamma$, where the
ratio of the sidelobe level to the main lobe is $0\leq\gamma\leq 1$,
for out-of-sector power which is both transmitted and received by
the sectorized antenna. Fig. \ref{fig:sectmodel} depicts the model.
The Table 1 conveys the power emitted by a transmitter in and out-of
sector subject to constant total power $\rho$. Under this model, the
following Proposition holds:

\begin{table}
\caption{Radiated Power Densities}
\begin{center}
\begin{tabular}{|c|c|c|c|}
\hline
--&In sector& Out of sector& Combined\\
\hline Power emitted:& $\frac{\rho}{1 + \gamma(M-1)}$& $\frac{\rho
\gamma(M-1)}{1+\gamma(M-1)}$& Sum: $\rho$\\
\hline Sector size (rad):& $2\pi\frac{1}{M}$& $
2\pi\frac{M-1}{M}$& Sum: $2\pi$\\
\hline Power density: &$\frac{M/2\pi}{1+\gamma(M-1)}$&$\frac{\gamma M/2\pi}{1+\gamma(M-1)}$& Ratio: $\gamma$\\
\hline
\end{tabular}
\end{center}
\end{table}

\paragraph*{Proposition 2} \emph{For a random access wireless network
in which nodes have $M$ sectorized directional antennas in
Nakagami-$m$ fading with a constant (fractional) sidelobe level
$\gamma\in[0,1]$ for out of sector power transmitted and received,
the optimal contention density with outage $\epsilon$ is given by:}
\begin{equation}
\bar{\lambda_\epsilon}=\left(\frac{M}{1+\gamma^{\frac{2}{\alpha}}(M-1)}\right)^2
\frac{K_{\alpha,m}\epsilon}{C_{\alpha,m}\beta^{\frac{2}{\alpha}}R^2}
\end{equation}
\emph{where $K_{\alpha,m}$ is given by (\ref{eq:kmrcP}) and
$C_{\alpha,m}$ by (\ref{eq:CP}). Thus $\gamma^{-\frac{4}{\alpha}}$
is an upper bound on the transmission capacity increase due to
antenna sectorization.}

\begin{proof}
We assume the fading statistics of the received signal are unchanged
by the sectorized antennas, but rather are merely scaled by the
emitted and received power density. That the fading statistics are
unchanged is reasonable for a small to moderate number of sectors,
while for very directional antennas, the scattering seen by any
given sector will be reduced and no longer have the typical
isotropic properties. As a result of sectorization, four
interference terms surface as follows: Let $\Phi_1$ be the set of
interferers which are in the active sector of the receiver of
interest and are transmitting toward the receiver. Let $\Phi_2$ be
the set of interferers in the receiver's active sector which are not
transmitting toward the receiver. Let $\Phi_3$ be the set of
interferers outside the receiver's sector which are transmitting
toward the receiver. And let $\Phi_4$ consist of those interferers
transmitting away from the receiver and which are not in the
receiver's sector. The independence property of the Poisson process
implies these four shot noise processes are independent as well.
Furthermore, the $I_{\Phi_i}$ are each related to $I_\Phi$ since
they occur over disjoint subsets of the plane (i.e., a certain
sector), are scaled by the combined antenna gains, and the point
process of interferers is thinned according to the direction the
interferers transmit. The table summarizes the interference
contributions from each of these processes with $\lambda_i$ the
effective node density of the process, $\theta_i$ the sector size
over which the process occurs from the perspective of the typical
receiver, and $\psi_i$ the combined antenna gains.

\begin{table}
\caption{Interference Shot Noise Processes}
\begin{center}
\begin{tabular}{|c|c|c|c|}
\hline
--&$\lambda_i$&$\theta_i$&$\psi_i$\\
\hline $I_{\Phi_1}$& $\lambda\frac{1}{M}$& $2\pi\frac{1}{M}$&$\left(\frac{M}{1+\gamma(M-1)}\right)^2$\\
\hline $I_{\Phi_2}$& $\lambda\frac{M-1}{M}$& $2\pi\frac{1}{M}$&$\gamma\left(\frac{M}{1+\gamma(M-1)}\right)^2$\\
\hline $I_{\Phi_3}$& $\lambda\frac{1}{M}$& $2\pi\frac{M-1}{M}$&$\gamma\left(\frac{M}{1+\gamma(M-1)}\right)^2$\\
\hline $I_{\Phi_4}$& $\lambda\frac{M-1}{M}$& $2\pi\frac{M-1}{M}$&$\gamma^2\left(\frac{M}{1+\gamma(M-1)}\right)^2$\\
\hline
\end{tabular}
\end{center}
\end{table}

The transforms of the shot noise processes of the $I_{\Phi_i}$ are
given by
\begin{equation}
\mathcal{L}_{\Phi_i}(\zeta)=\exp\left\{-\lambda_i\theta_i\int_0^{\infty}1-E\left[e^{-\zeta
\psi_i S|x|^{-\alpha}}\right]\mathrm{d}x\right\}
\end{equation}
for $\zeta=\psi^{-1}_0\beta R^{\alpha}$ with $\psi_0$ being the
combined antenna gain between the typical receiver and its intended
transmitter, and $\psi_0=\psi_1$. Consider the Rayleigh fading case.
The outage probability at a typical receiver is
\begin{eqnarray}
\mathbf{P}(SIR\geq\beta)&=&\mathbf{P}\left(\frac{\psi_0\rho S_0 R^{-\alpha}}
{\rho \sum_{i=1}^4 I_{\Phi_i}} \geq\beta \right)\nonumber\\
&=&\int_0^{\infty}F^c_{S_0}(\psi_0^{-1}\beta R^\alpha
s)f_{[I_{\Phi_1}+I_{\Phi_2}+I_{\Phi_3}+I_{\Phi_4}]}(s)\nonumber\\
&=&\prod_{i=1}^{4}\left.\mathcal{L}_{\Phi_i}(\zeta)\right|_{\zeta=\psi^{-1}_0\beta R^{\alpha}}\nonumber\\
&=&\exp\left\{-\lambda\beta^{\frac{2}{\alpha}}R^2 C_{\alpha}
\left(\frac{1+\gamma^{\frac{2}{\alpha}}(M-1)}{M}\right)^2\right\}\nonumber\\&&
\end{eqnarray}
and solving for $\lambda$ gives
\begin{equation}
\label{eq:lampfsectRay}
\bar{\lambda_\epsilon}=\left(\frac{M}{1+\gamma^{\frac{2}{\alpha}}(M-1)}\right)^2
\frac{\epsilon}{C_{\alpha}\beta^{\frac{2}{\alpha}}R^2}\cdot
\end{equation}

Next note that
$\frac{M}{1+\gamma^{\frac{2}{\alpha}}(M-1)}\leq\frac{1}{\gamma^{\frac{2}{\alpha}}}$
and as $M$ becomes large, we have the limit
\begin{equation}
\lim_{M\rightarrow
\infty}\left(\frac{M}{1+\gamma^{\frac{2}{\alpha}}(M-1)}\right)^2=\frac{1}{\gamma^{\frac{4}{\alpha}}}.
\end{equation}
This results in an upper bound of $\gamma^{-\frac{4}{\alpha}}$ on
the improvement (over (\ref{eq:naklamthm})) in optimal contention
density from sectorized antennas. If signals are Nakagami-$m$
distributed instead, since the desired and interfering signals are
independent, $C_{\alpha,m}$ replaces $C_\alpha$ and $K_{\alpha,m}$
appears in the numerator of (\ref{eq:lampfsectRay}).
\end{proof}

These results firstly indicate that directional antennas increase
transmission capacity by nearly a factor of $M^2$ for low sidelobe
levels. This indicates that MIMO techniques that avoid or reduce
interference in an ad hoc network are highly beneficial at the
physical layer. In addition there are advantages at higher network
layers such as increased ability to learn the topology of the
network, perform directional routing, etc; see \cite{IEEEbib:ra} and
\cite{IEEEbib:rc} and references therein for more details. This
section has characterized the potential increase in area spectral
efficiency due to antenna sectorization which by itself provides
greater potential and flexibility for routing and network
management, but the full relationship between directional antennas
and these higher layer functions is still an area of ongoing
research.

However, this analysis also indicates that if for practical reasons,
sidelobe levels cannot be reduced, then the sidelobes limit the
potential gains even for very directional antennas. This model also
suffers from very idealistic assumptions about the real propagation
environment, especially since dense multipath can result in signal
angle of arrival being quite different from the geographic angle to
the transmitter. As pointed out in \cite{IEEEbib:ra}, real antenna
patterns are far from ``pie slices'' and in multipath environments,
static antennas are much less robust to fluctuating channels.

\section{\label{sec:MRC}Transmission Capacity of Eigen-Beamforming Networks}

Dynamic beamforming is one of the most prominent multiple antenna
techniques, having been employed for decades in electromagnetic
detection and imaging applications. The complexity is manageable and
it can be performed on any number of antennas in any configuration
(\cite{IEEEbib:bal}, ch. 6). However, to be explicit since
``beamforming'' has become quite an overloaded term, this section
uses the term to mean the following: At the receiver it refers to a
coherent linear combination of the antenna outputs, while at the
transmitter it refers to sending linearly weighted versions of the
same signal on each antenna. Thus, unlike the previous section, no
attention is paid to the specific physical pattern of energy
propagation. In each case for this analysis, the weights are
determined by the dominant singular vectors or eigenvectors (hence,
``eigen-beamforming'') of the channel. Throughout this section it is
assumed that both the transmitter and receiver have perfect channel
knowledge of their own channel, but not of interfering channels.
Hence, signaling strategies will maximize SNR over a specific
channel but not necessarily SINR, though the analysis of the
resulting interference-limited systems will ultimately ignore
background thermal noise. We focus first on the vector (SIMO or
MISO) channel for which eigen-beamforming is equivalent to maximal
ratio transmission or combining. We then consider the general matrix
(MIMO) channel for which a single datastream is sent over the
dominant eigenmode.

\subsection{1 x M and M x 1 Eigen-Beamforming}

Consider first a wireless system in which all transmitters transmit
with power $\rho$ using only one antenna and receivers beamform on
$M$ antennas by coherently combining the received signals. Again,
this is beamforming along the dominant (and only) eigenmode of the
$1\times M$ channel. As shown in \cite{IEEEbib:ka}, this is
equivalent to an $M\times 1$ vector channel for which maximal ratio
transmission is performed at the transmitter and one receive antenna
is used. The channel model for the desired signal in a Rayleigh
fading environment is a vector of i.i.d. unit variance, complex
Gaussian entries scaled by the power law path loss function:
${\mathbf h}_0\sqrt{|R|^{-\alpha}}$ for the $k$th entry of ${\mathbf
h}_0$ independently $[\mathbf{h}_0]_k\sim \mathcal{CN}(0,1)$, and
similarly the channel between a receiver and the $i$th interferer is
${\mathbf h}_i\sqrt{|X_i|^{-\alpha}}$ with $[\mathbf{h}_i]_k\sim
\mathcal{CN}(0,1)$. Under this model, the following Proposition
holds. As in Sec. \ref{sec:nakag}, the Proposition will be given in
two parts: the first is an expression for the exact optimal
contention density for small outage constraints and the second is a
set of bounds that help interpret the exact results.
\paragraph*{Proposition 3} \emph{For a random access wireless network
in which nodes transmit on a single antenna and perform maximal
ratio combining with $M$ antennas; or equivalently perform maximal
ratio transmission with $M$ antennas and receive on a single
antenna; the optimal contention density under Rayleigh fading with
outage constraint $\epsilon$ is given by:}
\begin{equation}
\label{eq:lammrc}
\bar{\lambda_\epsilon}=\frac{K_{\alpha,M}\epsilon}{C_{\alpha}\beta^{\frac{2}{\alpha}}R^2}
\end{equation}
\emph{where $K_{\alpha,M}$ is given by (\ref{eq:kmrcP}) and
$C_{\alpha}=C_{\alpha,1}$ in (\ref{eq:CP}). Further,
$\bar{\lambda_\epsilon}$ is $\Theta(M^{\frac{2}{\alpha}})$ and
bounded by:}
\begin{equation}
\label{eq:boundsmrc}
\frac{M^{\frac{2}{\alpha}}\epsilon}{C_\alpha\beta^{\frac{2}{\alpha}}R^2}\leq
\bar{\lambda_\epsilon}\leq\frac{\Gamma(1-\frac{2}{\alpha})
M^{\frac{2}{\alpha}}\epsilon}{C_\alpha\beta^{\frac{2}{\alpha}}R^2}\cdot
\end{equation}

\begin{proof}
To characterize the interference seen by an $M$-antenna receiver
that ignores interfering signals, beamforming simply to maximize its
own received signal power (again thermal noise is assumed
negligible), the SIR expression is:
\begin{eqnarray}
SIR&=&\frac{\frac{\rho} {M}|{\mathbf h}^H_0{\mathbf
h}_0|^2R^{-\alpha}}{\frac{\rho}{M}\sum_{X_i\in
\Phi}|\mathbf{h}_0^H\mathbf{h}_i|^2|X_i|^{-\alpha}}\nonumber\\
&=&\frac{\|{\mathbf h}_0\|^2R^{-\alpha}}{\sum_{X_i\in \Phi}
\left|\frac{\mathbf{h}_0^H}{\|{\mathbf h}_0\|}{\mathbf h}_i
\right|^2|X_i|^{-\alpha}} \cdot
\end{eqnarray}
As shown in \cite{IEEEbib:sh}, since a linear combination of
Gaussian variables is again Gaussian, the product
$\frac{\mathbf{h}^H_0}{\|\mathbf{h}_0\|}\mathbf{h}_{i}$ is
distributed as a single complex Gaussian random variable with zero
mean and unit variance. Letting $S_i=\left|
\frac{\mathbf{h}^H_0}{\|\mathbf{h}_0\|}\mathbf{h}_{i}\right|^2$,
which is exponentially distributed, the SIR expression is
\begin{equation}
SIR=\frac{\|{\mathbf h}_0\|^2R^{-\alpha}}{\sum_{X_i\in \Phi}S_i
|X_i|^{-\alpha}}=\frac{\|{\mathbf h}_0\|^2R^{-\alpha}}{I_\Phi} \cdot
\end{equation}

Setting $S_0=\|{\mathbf h}_0\|^2$ and considering the network model
in Sec. \ref{sec:model} but with beamforming receivers with $M$
antennas, the distribution of the received signal is now $\chi^2$
with $2M$ degrees of freedom. The CCDF of $S_0$ is
$F^c_{S_0}(x)=e^{-x}\sum_{k=0}^{M-1}\frac{x^k}{k!}$. However, the
interference has the same form as the shot noise process for the
single-antenna case. If we now apply a small outage constraint and
Theorem 1, we can state simply that $K_{\alpha,M}$ is given by
(\ref{eq:kmrcP}) and $C_\alpha=C_{\alpha,1}$ in (\ref{eq:CP}). As
shown in (\ref{eq:Kbound}),
\begin{equation*}
1\leq\frac{K_{\alpha,M}}{M^{\frac{2}{\alpha}}}\leq
\Gamma(1-2/\alpha)
\end{equation*}
which indicates that
\begin{eqnarray}
\frac{M^{\frac{2}{\alpha}}\epsilon}{C_\alpha\beta^{\frac{2}{\alpha}}R^2}&\leq&
\frac{K_{\alpha,M}\epsilon}{C_\alpha\beta^{\frac{2}{\alpha}}R^2}
\leq\;\;\frac{\Gamma(1-\frac{2}{\alpha})M^{\frac{2}{\alpha}}\epsilon}{C_\alpha\beta^{\frac{2}{\alpha}}R^2}
\end{eqnarray}
with equality to the lower bound at $M=1$ and approaching the upper
bound with increasing $M$ since $C_\alpha$ is constant while
$K_{\alpha,M}$ is increasing in $M$. The term in the middle is now
equal to $\bar{\lambda_\epsilon}$.
\end{proof}

\begin{figure}
\begin{center}
\resizebox{9cm}{7cm}{\includegraphics{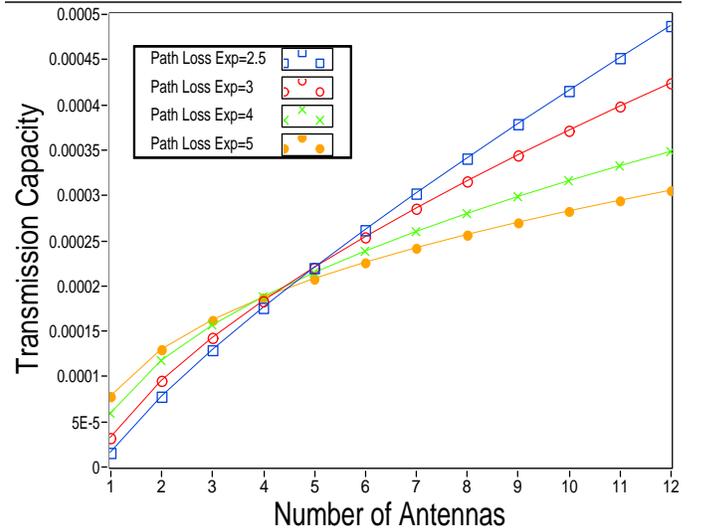}}
\caption{Transmission capacity versus $M$ for four path loss
exponents for $1\times M$ MRC. Higher path loss separates
transmissions spatially and is the dominant effect for smaller
numbers of antennas. But with a larger number of antennas,
ultimately network performance is improved more through interference
robustness than spatial separation.} \label{fig:MRCalpha2}
\end{center}
\end{figure}

\begin{figure}
\begin{center}
\resizebox{9cm}{7cm}{\includegraphics{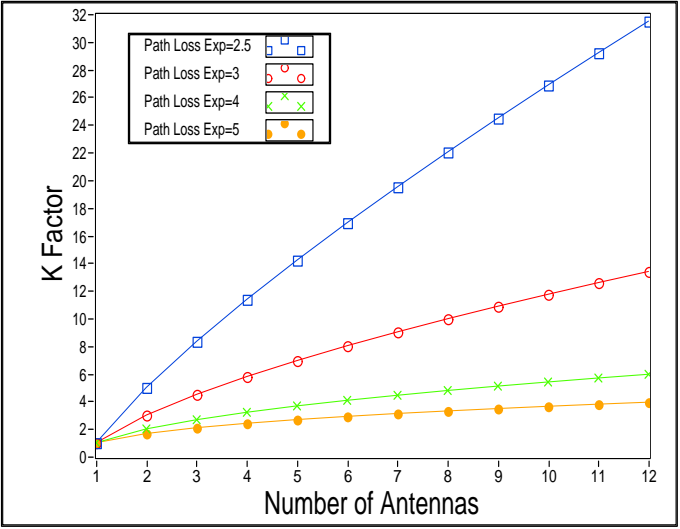}}
\caption{The $K_{\alpha,M}$ factor versus $M$ for four path loss
exponents for $1\times M$ MRC. Lower path loss results in much
greater gains over the SISO case ($M=1$).} \label{fig:MRCalpha}
\end{center}
\end{figure}

\begin{figure}
\begin{center}
\resizebox{9cm}{7cm}{\includegraphics{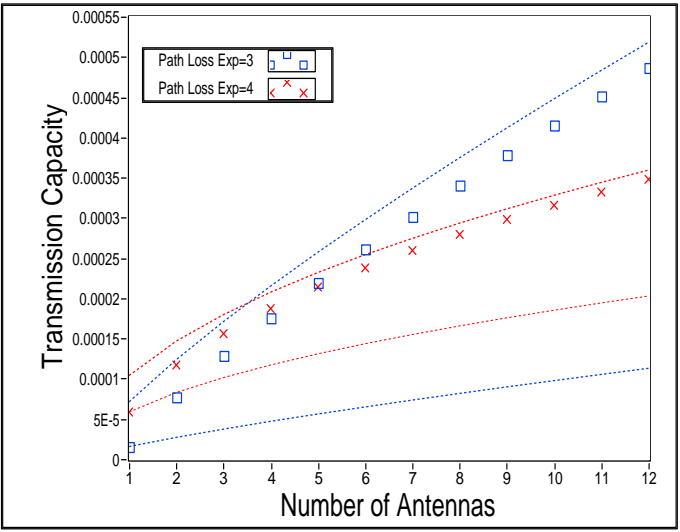}}
\caption{Demonstration of the bounds on transmission capacity for
$1\times M$ or equivalently $M\times 1$ MRC. The upper bound is
asymptotically tight and a good approximation for higher path loss.}
\label{fig:MRCbounds}
\end{center}
\end{figure}

Proposition 3 gives a general scaling of the optimal contention
density with the number of antennas, target SIR, path loss, the
transmitter-receiver separation, and the outage constraint. Fig.
\ref{fig:MRCalpha2} gives the transmission capacity versus $M$ for
four different path loss exponents. Fig. \ref{fig:MRCalpha} gives
the $K_{\alpha,M}$ factor versus $M$ for the same path loss
exponents. As evident from the figures, as path loss reduces and
interference becomes less attenuated by distance, the gain of the
MIMO technique over the SISO case increases. However, higher path
loss results in higher transmission capacity for smaller numbers of
antennas since path loss helps to spatially separate transmissions.
Fig. \ref{fig:MRCbounds} demonstrates the relationship of the exact
$K_{\alpha,M}$ factor to the upper and lower bounds. The upper bound
is both asymptotically tight and a good approximation for higher
path loss.

\subsection{\label{sec:MRT} $M_t\times M_r$ MIMO Eigen-Beamforming}

Now consider the same network but with nodes each equipped with
$M_t$ transmit and $M_r$ receive antennas to perform dynamic
eigen-beamforming at both transmitter and receiver ends. This
extension of MRC has significant advantages even over $1\times M$
MRC since the diversity order increases as $M_tM_r$. The Rayleigh
fading MIMO channel is modeled as a matrix of i.i.d. zero-mean,
unit-variance complex Gaussian entries scaled by path loss. The
channel of the desired signal for the transmitter-receiver pair of
interest is denoted ${\mathbf H}_{00}$. The transmitter and receiver
beamform using the input and output singular vectors ${\mathbf v}_0$
and ${\mathbf u}_0$, respectively, corresponding to the maximum
singular value of ${\mathbf H}_{00}$. This results in the received
power being equal to the square of the maximum singular value
$\phi^2_{\mathrm{max}}$ scaled by path loss and the transmit power.
Each interfering transmitter, on the other hand, beamforms to
maximize received power across some other Rayleigh channel ${\mathbf
H}_{ii}$ using beamforming vector ${\mathbf v}_i$, and interferes at
the receiver of interest through channel ${\mathbf H}_{0i}$. For
such a network, the following bounds hold:
\paragraph*{Proposition 4} \emph{For a random access wireless network in which
nodes perform maximal ratio transmission and combining on $M_t$ and
$M_r$ antennas respectively, for small outages the optimal
contention density is bounded by:
\begin{equation}
\frac{\max\{M_t,M_r\}^{\frac{2}{\alpha}} \epsilon}{C_\alpha
R^2\beta^{\frac{2}{\alpha}}}\leq\bar{\lambda_\epsilon}\leq\frac{\Gamma(1-\frac{2}{\alpha})(M_tM_r)^{\frac{2}{\alpha}}
\epsilon}{C_\alpha R^2\beta^{\frac{2}{\alpha}}}
\end{equation}
for $C_\alpha=C_{\alpha,1}$ given in (\ref{eq:CP}).}

\begin{proof}
To begin, the SIR expression for this model is
\begin{equation}
\label{eq:mrtsinr} SIR=\frac{\rho\phi^2_{\mathrm{max}}R^{-\alpha}}
{\rho\sum_{X_i\in \Phi}|X_i|^{-\alpha}
|\mathbf{u}^H_0\mathbf{H}_{0i}\mathbf{v}_i|^2}\:\cdot
\end{equation}
Note that $\mathbf{u}_0$, $\mathbf{H}_{0i}$, and $\mathbf{v}_i$ are
all independent. As discussed in \cite{IEEEbib:ka2}, the full
product $\mathbf{u}^H_0\mathbf{H}_{0i}\mathbf{v}_i$ is distributed
as a single zero-mean, unit-variance, complex Gaussian variable
since the inner product of a vector i.i.d. complex Gaussian
variables with an arbitrary unit vector is a single complex Gaussian
variable. This simplifies the SIR expression to
$SIR=\frac{\phi^2_{\mathrm{max}}R^{-\alpha}}{I_\Phi}$. with the
distribution of the interference unchanged from the single antenna
Rayleigh fading case. Again neglecting thermal noise, the received
and interfering signals are independent and
$C_{\alpha}=C_{\alpha,1}$ in (\ref{eq:CP}) by equivalence of the
shot noise processes.

As for the received signal, note that the CCDF of the square of the
maximum singular value of the desired channel (or equivalently the
largest eigenvalue of a complex Wishart matrix), has been reported
by Kang and Alouini \cite{IEEEbib:ka} (originally given by Khatri
\cite{IEEEbib:kh}):
\begin{equation*}
F^c_{\phi^2_{\mathrm{max}}}(x)=1-\frac{|\Psi(x)|}{\Pi^q_{k=1}\Gamma(q-k+1)\Gamma(s-k+1)}
\end{equation*}
where $|\cdot|$ denotes a determinant, $q=\min\{M_t,M_r\}$,
$s=\max\{M_t,M_r\}$, and the entries of the $q\times q$ matrix
$\Psi(x)$ are given by
$\{\Psi(x)\}_{i,j}=\gamma(s-q+i+j-1,x)\;,\;i,j=1,\;...\;,q$ where
$\gamma(\cdot,\cdot)$ is the lower incomplete gamma function. Recall
$\gamma(n,x)=(n-1)!
\left(1-e^{-x}\sum_{k=0}^{n-1}\frac{x^k}{k!}\right)$, for $n\in
\mathbb{N}$. This now facilitates the application of Theorem 1
yielding the outage probability and the optimal contention density,
which will again have the form:
\begin{equation*}
\bar{\lambda_\epsilon}=\frac{K^{mrt}_{\alpha,M_t,M_r}
\epsilon}{C_\alpha R^2\beta^{\frac{2}{\alpha}}}\: \cdot
\end{equation*}
We are unable to give an expression for $K^{mrt}_{\alpha,M_t,M_r}$
since the explicit sum-of-exponentials-and-polynomials form for
$F^c_{\phi^2_{\mathrm{max}}}$ is not known. However, the largest
squared singular value is bounded by \cite{IEEEbib:hj}:
\begin{equation*}
\|\mathbf{H}_{00}\|^2_F\geq
\phi^2_{\mathrm{max}}\geq\frac{\|\mathbf{H}_{00}\|^2_F}{
\min\{M_t,M_r\} }.
\end{equation*}
Since $\|\mathbf{H}_{00}\|^2_F$ is $\chi^2$ with $2M_tM_r$ degrees
of freedom this is equivalent to a particular MRC case in
(\ref{eq:lammrc}) and (\ref{eq:boundsmrc}) indicating that
\begin{equation}
K^{mrt}_{\alpha,M_t,M_r}\leq K_{\alpha,(M_tM_r)}\leq
\Gamma(1-2/\alpha)(M_tM_r)^{\frac{2}{\alpha}}
\end{equation}
so that
\begin{equation}
\bar{\lambda_\epsilon}\leq
\frac{\Gamma(1-\frac{2}{\alpha})(M_tM_r)^{\frac{2}{\alpha}}
\epsilon}{C_\alpha R^2\beta^{\frac{2}{\alpha}}}\cdot
\end{equation}
Furthermore, a lower bound can be obtained from (\ref{eq:boundsmrc})
as
\begin{eqnarray}
K^{mrt}_{\alpha,M_t,M_r}&\geq&
\frac{K_{\alpha,(M_tM_r)}}{\min\{M_t,M_r\}^{\frac{2}{\alpha}}}\nonumber\\
&\geq&\frac{(M_tM_r)^{\frac{2}{\alpha}}}{\min\{M_t,M_r\}^{\frac{2}{\alpha}}}\nonumber\\
&=&\max\{M_t,M_r\}^{\frac{2}{\alpha}}
\end{eqnarray}
so that
\begin{equation}
\bar{\lambda_\epsilon}\geq\frac{\max\{M_t,M_r\}^{\frac{2}{\alpha}}\epsilon}{C_\alpha
R^2\beta^{\frac{2}{\alpha}}}\:\cdot
\end{equation}
\end{proof}

\begin{figure}
\begin{center}
\resizebox{9cm}{7cm}{\includegraphics{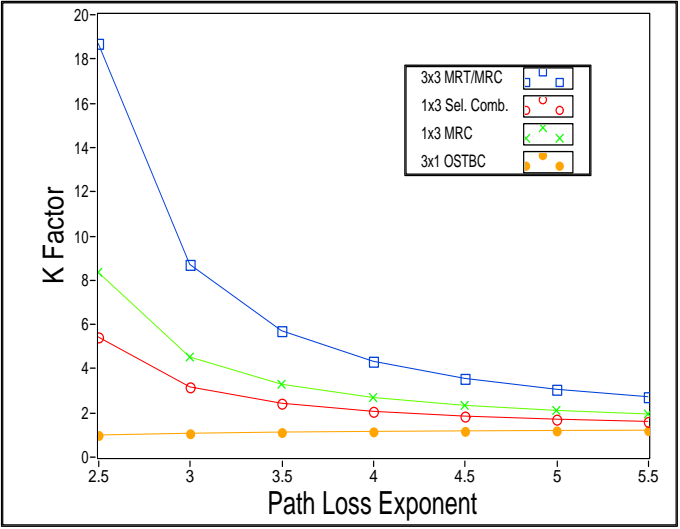}} \caption{
$K^{mrt}_{\alpha,3}$ ($3\times 3$ MRT/MRC), $K_{\alpha,3}$ ($1\times
3$ MRC), $K^{sc}_{\alpha,M}$ ($1\times 3$ selection combining), and
$\frac{K_{\alpha,M_t}}{C_{\alpha,M_t}}C_{\alpha,1}$ ($3\times 1$
OSTBC) versus $\alpha$ for various $M$. The factor for OSTBCs is the
only which increases (slightly) with increasing path loss since
interference reduces. (The factor $C_{\alpha,1}$ is a normalizing
factor for fair comparison with other techniques and the SISO case
for which $K=1$.)} \label{fig:labAlpha}
\end{center}
\end{figure}

Since $K^{mrt}_{\alpha,M_t,M_r}$ cannot be given explicitly for
arbitrary $M_t$ and $M_r$ at present, consider as an example the
case $M_t=M_r=2$ for which
\begin{equation*}
F^c_{\phi^2_{\mathrm{max}}}(x)=2e^{-x}-e^{-2x}+x^2e^{-x}.
\end{equation*}
Applying Theorem 1 for small outages:
\begin{equation}
\mathbf{P}(SIR<\beta)\approx\lambda C_\alpha
R^2\beta^{\frac{2}{\alpha}}
\left(2-2^{\frac{2}{\alpha}}-\frac{2}{\alpha}\left(1-\frac{2}{\alpha}\right)\right).
\end{equation}
For $M_t=M_r$ and both large, $\phi^2_{\mathrm{max}}$ of the channel
matrix approaches $4M_t$ \cite{IEEEbib:ae}. This leads to the
conjecture that for moderately large numbers of antennas (e.g.,
$M_t,M_r>3$), the lower bound reflects the orderwise behavior in a
rich scattering environment. However, the upper bound should be more
appropriate in a LOS channel. Fig. \ref{fig:labAlpha} depicts both
$K_{\alpha,M}$ and for the square channel case $M=M_t=M_r$,
$K^{mrt}_{\alpha,M}$ for various $M$, $\alpha$. Again
$K^{mrt}_{\alpha,M}\rightarrow 1$ with increasing $\alpha$. This
implies that as $\alpha$ becomes large, nodes are already spatially
separated through path loss, and spatial diversity yields less
improvement over the single antenna case.

We also point out that the expression for
$F^c_{\phi^2_{\mathrm{max}}}$ in Rayleigh fading channels, repeated
above from \cite{IEEEbib:ka}, is given for any number of antennas at
either the transmitter or receiver. The result is always a sum of
terms of the form $x^ke^{-nx}$ so that the Laplace transform method
used here may be applied for systems with any number of antennas.
For larger numbers of antennas especially, there also remains the
question if spatial multiplexing has a place. While this is left for
future investigation, it should be noted that the form of the joint
distribution of the eigenvalues of Wishart matrices given in
\cite{IEEEbib:te} indicates that the Laplace transform method can be
extended to the spatial multiplexing case.

\section{\label{sec:ostbc}Transmission Capacity of OSTBC Networks}

Orthogonal space-time block coding has been one of the more quickly
accepted transmit diversity techniques for several good reasons.
First, OSTBCs achieve full diversity in point-to-point links without
requiring channel state information at the transmitter. Second, an
optimum receiver design is simply a matched filter without any need
for joint decoding of multiple symbols (error correction codes
notwithstanding). Furthermore, space-time coding results in far less
variability in the effective channel, greatly reducing the frequency
and duration of deep fades. However, there is another source of
effective channel instability, particularly in decentralized
networks, which is cochannel interference. In light of the results
on reduced fading as well as MRT/MRC earlier in this paper, for
which the latter results in much greater network improvement, it is
unclear how OSTBCs compare in a decentralized, interference-limited
environment and warrants further investigation.

Specific codes are characterized by the number of transmit antennas
used ($M_t$), the number of time slots used ($N$), and the number of
independent data symbols sent ($N_s$) \cite{IEEEbib:ls}. Again $M_r$
denotes the number of receive antennas but this has no effect on the
code structure. However, it will also be necessary to characterize
OSTBCs by the number of time slots over which each symbol is
repeated ($N_r$). The familiar Alamouti code has $M_t=N=N_s=N_r=2$.
\paragraph*{Proposition 5} \emph{For a random access wireless network
in which transmitting nodes use orthogonal space-time block codes
with $M_t$ transmit antennas and code parameter $N_r$ and receiving
nodes perform maximal ratio combining with $M_r$ antennas in
Rayleigh fading, the optimal contention density under the outage
constraint $\epsilon$ is given by:}
\begin{equation}
\bar{\lambda_\epsilon}=\frac{K_{\alpha,M_tM_r}\epsilon}{C_{\alpha,N_r}\beta^{\frac{2}{\alpha}}R^2}
\end{equation}
\emph{where $K_{\alpha,M_tM_r}$ is given by (\ref{eq:kmrcP}) and
$C_{\alpha,N_r}$ in (\ref{eq:CP}). Further, $\bar{\lambda_\epsilon}$
is $\Theta(M_r^{\frac{2}{\alpha}})$:}
\begin{equation}
\frac{M_r^{\frac{2}{\alpha}}\epsilon}{C_{\alpha,1}\beta^{\frac{2}{\alpha}}R^2}\leq
\bar{\lambda_\epsilon}\leq\frac{M_r^{\frac{2}{\alpha}}\epsilon}{\pi\beta^{\frac{2}{\alpha}}R^2}\cdot
\end{equation}

\begin{proof}
For the received signal, since detection decouples for OSTBCs
\cite{IEEEbib:pj} over the $N_r$ time slots as well as $M_r$
antennas, the received amplitude is $\|\mathbf{H}_0\|^2_F$ for each
symbol \cite{IEEEbib:pj}, where $\mathbf{H}_0$ is the $M_r\times
M_t$ complex Gaussian channel. The distribution of
$\|\mathbf{H}_0\|^2_F$ is $\chi^2$, just as with MRC, but with $2M_t
M_r$ degrees of freedom. So applying Theorem 1, the $K$ factor is
$K_{\alpha,M}$ in (\ref{eq:kmrcP}) for $M=M_tM_r$.

The interference seen by an OSTBC processing system is more
complicated, however. To determine the distribution of $S_i$,
consider the expression for the interference term from a single
interferer:
\begin{equation}
|X_i|^{-\alpha}S_i=|X_i|^{-\alpha}\sum^{N_r}_{k=1}
\frac{\mathbf{h}_0^{H}}{\|\mathbf{h}_0\|}\mathbf{h}^{(k)}_i
s^{(k)}_i=|X_i|^{-\alpha}\sum^{N_r}_{k=1}S^{(k)}_i
\end{equation}
where $\mathbf{h}_0=\mathrm{vec}(\mathbf{H}_{0})$ and
$\mathbf{h}^{(k)}_i$ is a permutation of the entries in
$\mathrm{vec}(\mathbf{H}_i)$ depending on the block coding
structure. Since desired symbols are repeated $N_r$ times, each
$S_i$ is a sum $N_r$ terms $S^{(k)}_i$ each of which is
exponentially distributed though not independent. In a strict sense,
this violates the independence of $S_0$ and $S_i$ required by
Theorem 1. However, we assume rough independence of received and
interfering signal statistics, with the statistics of the sum of
$S^{(k)}_i$ nearly indistinguishable from a Gamma distribution
independent of $S_0$. The nature of the post-processing interference
here was also reported in \cite{IEEEbib:wc}. Note that this
assumption removes some inherent structure in the interference so
that the analysis becomes worst case.

Since the Gamma distribution is the same mark distribution
encountered for Nakagami-$m$ fading interferers, $C_{\alpha,N_r}$ is
given by (\ref{eq:CP}). As shown before, the factor increases with
$N_r^{\frac{2}{\alpha}}$ indicating that repeating the symbols
introduces more cochannel interference. Applying Theorem 1 for small
outages
\begin{equation}
\bar{\lambda_\epsilon}
=\frac{K_{\alpha,M_tM_r}\epsilon}{C_{\alpha,N_r}R^2\beta^{\frac{2}{\alpha}}}
\leq \frac{(M_tM_r)^{\frac{2}{\alpha}}\epsilon}{\pi
N_r^{\frac{2}{\alpha}}R^2\beta^{\frac{2}{\alpha}}}=\frac{M_r^{\frac{2}{\alpha}}\epsilon}{\pi
R^2\beta^{\frac{2}{\alpha}}}
\end{equation}
for most practical block codes since $M_t=N_r$, which is the best
case. For the lower bound, we simply ignore the change in the
constant $C_{\alpha,N_r}$ substituting $C_{\alpha,1}$ which is
greater than $\pi$. (That is, let $K_{\alpha,M}$ increase but not
$C_{\alpha,N_r}$).
\end{proof}

The primary insight from the analysis of OSTBCs is that in an
environment of significant cochannel interference, they accomplish
little. As is evident from the bounds, the number of receive
antennas is the primary factor in network performance. While block
codes harden the channel resulting in a network performance gain
equivalent to that gained from reducing fading, they also tend to
amplify interference since symbols are repeated. When symbols are
repeated multiple times from the same antenna, as in some orthogonal
designs, this effect is worsened so that $N_r=M_t$ is the best case.
Furthermore, even though power is split between simultaneously
transmitted symbols, the transmit antennas become multiple
independent interference sources for other nodes in the network.
Furthermore, OSTBCs take a hit in the data rate for any code beside
Alamouti's. So for a larger number of antennas, OSTBCs are typically
inferior to other schemes and are likely not worth even the slight
added complexity.

\begin{figure}
\begin{center}
\resizebox{9cm}{7cm}{\includegraphics{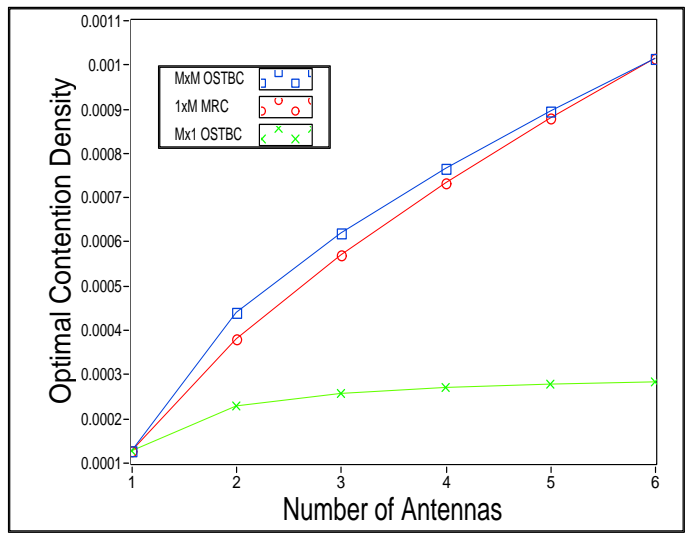}}
\caption{Comparison of the optimal contention density for $M\times
M$ OSTBC/MRC, $1\times M$ MRC, and $M\times 1$ OSTBC respectively.
The optimal contention density is nearly identical to receiver MRC
only and does not improve much for large numbers of transmit
antennas over the single antenna case without receive beamforming.}
\label{fig:OSTBCcomp}
\end{center}
\end{figure}

Fig. \ref{fig:OSTBCcomp} compares the optimal contention density for
$M\times 1$ OSTBCs for which the receiver receives on only one
antenna, $M\times M$ OSTBCs for which the receiver performs MRC on
$M$ antennas in addition to the transmit block coding, as well as
$1\times M$ MRC without block coding. The figure shows the optimal
contention density for $\alpha=3$, target SINR 4.77dB, and
transmitter-receiver separation 10m. First, there is little gain
over simply performing MRC in contention density. But what is not
shown is that for the number of antennas larger than two, the
transmission capacity for $M\times M$ OSTBCs actually falls below
the MRC curve since it must use a reduced rate code. If only one
receive antenna is used, then for any number of transmit antennas
beyond two, there is essentially no gain when code rate is taken
into account. This confirms that the primary source of gain is at
the receiver and that for any system beyond $2\times 2$, it would be
better to simply select one antenna and operate in the $1\times M$
MRC mode.

\section{\label{sec:selcomb}Transmission Capacity of Selection Diversity and Combining Networks}

A fundamental characteristic of MIMO fading channels is that due to
polarization, pattern diversity, or spatial separation, one or more
antenna elements may be receiving above average signal strength.
Simply selecting the best often has the practical advantage over
more sophisticated combining schemes of simpler implementation, or
less expensive hardware. There are a variety of ways to perform
antenna selection, and antenna selection can be used in conjunction
with other diversity techniques. As an example let the transmitter
operate one antenna and the receiver select one of $M$ which has the
best instantaneous channel with i.i.d. Rayleigh fading between all
antennas. In this case,
\begin{equation}
\label{eq:selcomb1xM}
F^c_{S_0}(x)=1-(1-e^{-x})^M=\sum_{k=1}^{M}{M\choose k}(-1)^{k+1}
e^{-kx}
\end{equation}
and the interference fading channels $S_i$ remain exponentially
distributed. This can be extended by considering a system that
selects the best pair of antennas (one transmit and one receive)
from among $M_t$ transmit and $M_r$ receive antennas. The parameter
$M$ in (\ref{eq:selcomb1xM}) is simply replaced by $M_tM_r$. Here
the full matrix channel is $\mathbf{H}_{00}$ as in Sec.
\ref{sec:MRT} from which the element with the largest magnitude is
selected. The following Proposition characterizes the gain from
selection diversity:
\paragraph*{Proposition 6} \emph{For a random access wireless network
in which nodes perform selection diversity/combining by selecting
the best pair among $M_t$ transmit and $M_r$ receive antennas in
Rayleigh fading, the optimal contention density under the outage
constraint $\epsilon$ is given by:
\begin{equation}
\bar{\lambda_\epsilon}=\frac{K^{sc}_{\alpha,M^2}}{C_\alpha}\frac{\epsilon}{\beta^{\frac{2}{\alpha}}R^2}
\end{equation}
for $M^2=M_tM_r$, $C_\alpha=C_{\alpha,1}$ in (\ref{eq:CP}), and}
\begin{equation}
K^{sc}_{\alpha,M^2}=\left[\sum_{k=1}^{M^2}{M^2\choose k}(-1)^{k+1}
k^{\frac{2}{\alpha}}\right]^{-1}.
\end{equation}

\begin{proof}
This is given by simply substituting the coefficients in
(\ref{eq:selcomb1xM}) into Theorem 1 and noting that the statistics
of the interference are identical to the SISO case for any pair of
antennas.
\end{proof}

\begin{figure}
\begin{center}
\resizebox{9cm}{7cm}{\includegraphics{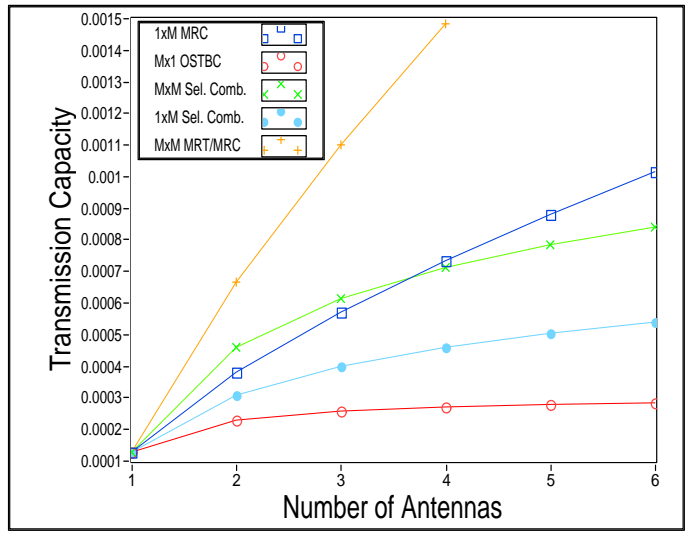}}
\caption{Comparison of the transmission capacity for MRT, MRC,
OSTBC, and selection combining. Note that $M\times M$ OSTBC/MRC is
nearly identical to the $1\times M$ MRC case. Also note that
$M\times M$ selection combining means selecting the best pair of
transmitting and receiving antennas (one of
each).}\label{fig:labcomp}
\end{center}
\end{figure}

There are a number of other distributions resulting from antenna
selection that can be considered. For example, in an $M_r\times M_t$
system performing MRC at the receiver, one transmit antenna may be
selected which has the largest magnitude vector channel to the
intended receiver. The distribution of the interference after MRC
processing will remain the same but $S_0$ will have
\begin{eqnarray}
\label{eq:selcombMtxMr}
F^c_{S_0}(x)&=&1-\left(1-e^{-x}\sum_{k=0}^{M_r-1}\frac{x^k}{k!}\right)^{M_t}\nonumber\\
&=&\sum_{m=1}^{M_t}e^{-mx}\sum_{k=0}^{M_t(M_r-1)}a_{mk}x^k
\end{eqnarray}
for
\begin{equation}
a_{mk}=(-1)^{M_t+m}{M_t\choose m}\sum_{\substack{n_1,n_2,...n_m\leq
M_r-1\\n_1+n_2...+n_{m}=k}}\prod_{i=1}^{m}(n_i!)^{-1}
\end{equation}
with the sum running over all (ordered) $m$-tuples of positive
integers less than $M_r-1$ which add to $k$. From Theorem 1 the $K$
factor can now be determined which specifies the optimal contention
density as well. Fig. \ref{fig:labcomp} compares the gain in
transmission capacity for a number of systems versus the number of
antennas, including MRT/MRC, OSTBCs, as well as two kinds of
selection diversity/combining: one in which the transmitter
transmits on one antenna and the receiver selects the best of its
own antennas, and the second in which the receiver and transmitter
jointly select the best pair of single antennas. Clearly antenna
selection can significantly enhance network performance since it
improves the typical channel without amplifying interference. Of
course, as the number of antennas becomes large, obtaining $M^2$
statistically independent pairs is difficult, and array gain quickly
becomes superior in terms of network performance. Still, Proposition
6 implies that antenna selection may be a desirable tradeoff in
terms of performance and complexity.

\section{\label{sec:concl}Conclusion}

In this paper, the performance of random access ad hoc networks
employing a number of spatial diversity techniques was determined.
Exact outage probabilities were derived for random wireless networks
as well as optimal contention densities for small outage constraints
for a large class of received signal distributions. These
distributions include those applicable for nodes employing maximal
ratio transmission/combining, orthogonal space-time block coding,
selection diversity/combining, and static beamforming with
sectorized antennas. The improvement in transmission capacity for
Nakagami-$m$ fading channels in which fading is reduced were also
given and shown to be equivalent to the small gains due to
space-time codes. The results show a significant improvement in
transmission capacity for sectorized antennas and beamforming
systems, a lesser but still appreciable gain for selection combining
systems, and marginal gains at best for space-time block coded
systems. Gains are higher for beamforming and selection combining
systems when interference is more severe both when node density
increases and under lower path loss, while the opposite is true for
space-time block coded systems. In general it was found that
diversity techniques employed at the receiver offer the most
practical benefits. Future research should address the enhancements
achievable from spatial multiplexing, multiuser MIMO techniques, and
the combined effect of MIMO at the physical layer for scheduling,
routing, and network management applied to ad hoc networks.

\appendices
\section{\label{sec:pflem1}Proof of Theorem 1}

\noindent Define the PDF of $I_\Phi$ to be
$f_{I_\Phi}(t)=\mathrm{d}\mathbf{P}(I_\Phi\leq t)$. Define a
transform of $f_{I_\Phi}(t)$ using CCDF $F^c_{S}(t)$ as
\begin{equation*}
\mathcal{G}_{I,S}(s)=\int_0^{\infty}F^c_{S}(st)f_{I_\Phi}(t)\mathrm{d}t.
\end{equation*}
The probability of successful transmission can be expressed as
\begin{eqnarray}
\mathbf{P}(SIR\geq\beta)&=&\mathbf{P}(S\geq \beta
R^{\alpha}I_\Phi)\nonumber\\ &=&
\int_0^{\infty}F^c_{S}(st)f_{I_\Phi}(t)\mathrm{d}t\nonumber\\
&=& \left.\mathcal{G}_{I,S}(s)\right|_{s=\beta R^{\alpha}}
\end{eqnarray}
When $S\sim\mathrm{Exp}(1)$, $F^c_S(t)=e^{-t}$ so that the transform
of $f_{I_\Phi}(t)$ is
\begin{eqnarray}
\mathcal{G}_{I,S}(s)&=&\int_0^{\infty}F^c_{S}(st)f_{I_\Phi}(t)\mathrm{d}t\nonumber\\
&=& \mathcal{L}\{f_{I_\Phi}(t)\}(s)=\mathcal{L}_{I_\Phi}(s),
\end{eqnarray}
and the transmission success probability is expressible in terms of
the Laplace transform. Next suppose $F^c_S(t)=\sum_n e^{-nt}\sum_k
a_{nk}t^k,$ then the transform of $f_{I_\Phi}(t)$ using CCDF
$F^c_{S}(t)$ is
\begin{eqnarray}
\mathcal{G}_{I,S}(s)&=&\int_0^{\infty}F^c_{S}(st)f_{I_\Phi}(t)\mathrm{d}t
\nonumber\\ &=&\int_0^{\infty}\left(\sum_n e^{-nst}\sum_k
a_{nk}(st)^k\right)f_{I_\Phi}(t)\mathrm{d}t\\
&=&\sum_n \sum_k a_{nk}s^k\left(\int_0^{\infty}e^{-nt}t^k
f_{I_\Phi}(t)\mathrm{d}t\right)\\
&=&\sum_n\sum_k a_{nk}s^k\mathcal{L}\{t^k f_{I_\Phi}(t)\}(ns)\label{eq:lapapp}\\
&=&\sum_n\sum_k
a_{nk}(-s)^k\frac{\mathrm{d}^k}{\mathrm{d}(ns)^k}\mathcal{L}_{I_\Phi}(ns)\\
&=&\sum_n\sum_k\left[
a_{nk}\left(-\frac{\zeta}{n}\right)^k\frac{\mathrm{d}^k}{\mathrm{d}\zeta^k}\mathcal{L}_{I_\Phi}(\zeta)
\right]_{\zeta=n\beta R^\alpha}
\end{eqnarray}
where (\ref{eq:lapapp}) uses the Laplace transform property $
t^nf(t)\longleftrightarrow
(-1)^n\frac{\mathrm{d}^n}{\mathrm{d}s^n}\mathcal{L}[f(t)](s)$. To
derive (\ref{eq:K}), the derivatives of $\mathcal{L}_{I_\Phi}$ are
needed and they are given by
\begin{eqnarray}
\label{eq:deriv}
\frac{\mathrm{d}^p}{\mathrm{d}\zeta^p}\mathcal{L}_{I_\Phi}(\zeta)=\frac{e^{-\lambda
\zeta^{\frac{2}{\alpha}}C_\alpha}}{(-\zeta)^p}\sum_{k=1}^{p}\left(\lambda
\zeta^{\frac{2}{\alpha}}C_\alpha\frac{2}{\alpha}\right)^k(-1)^{k}\Upsilon_{p,k}
\end{eqnarray}
where
\begin{eqnarray*}
\Upsilon_{p,k}&=&\sum_{\delta_j\in\:\mathrm{comb}{p-1\choose
p-k}}\:\prod_{l_{ij}\in\delta_j}\left(\frac{2}{\alpha}(l_{ij}-i+1)-l_{ij}\right),\nonumber\\
&& i=1,2,...,|\delta_j|,\:j=1,2,...,{p-1\choose p-k}.
\end{eqnarray*}
Here we define $\mathrm{comb}{a\choose b}$ as the set of all subsets
of the natural numbers $\{1,2,...,a\}$ of cardinality $b$ with
distinct elements, i.e., $\mathrm{comb}{a\choose b}$ is the set of
combinations of $\{1,2,...,a\}$ taken $b$ at a time. Thus there are
${a\choose b}$ subsets in $\mathrm{comb}{a\choose b}$ each with $b$
elements and the $\delta_j$ each constitute one such subset.

Forming the first order Taylor expansion for the $p$th derivative
around $\kappa=\lambda\beta^{\frac{2}{\alpha}}R^2C_\alpha=0$, note
that any term with $\kappa^k$ for $k>1$ is $o(\kappa)$ and can be
discarded so that
$\left.\frac{\mathrm{d}^p}{\mathrm{d}\zeta^p}\mathcal{L}_{I_\Phi}(\zeta)\right|_{\zeta=\beta
R^\alpha}$ reduces to
\begin{eqnarray}
\frac{\mathrm{d}^p}{\mathrm{d}\zeta^p}\mathcal{L}_{I_\Phi}(\zeta)&=&(-1)^{p+1}
\frac{e^{-\lambda \zeta^{\frac{2}{\alpha}}C_\alpha}}{\zeta^p}\lambda
\zeta^{\frac{2}{\alpha}}C_\alpha
(2/\alpha)\Upsilon_{p,1}\nonumber\\
&&+\Theta(\kappa^2)\nonumber\\
&=&(-1)^p e^{-\lambda \zeta^{\frac{2}{\alpha}}C_\alpha}\lambda
\zeta^{\frac{2}{\alpha}-p}C_\alpha
\prod_{l=0}^{p-1}(l-2/\alpha)\nonumber\\
&&+\Theta(\kappa^2)\nonumber\\
&=&(-1)^p\lambda \zeta^{\frac{2}{\alpha}-p}C_\alpha
\prod_{l=0}^{p-1}(l-2/\alpha)+\Theta(\kappa^2)\nonumber\\
&& \label{eq:TayBegin}
\end{eqnarray}
where the small error terms are the result of the Taylor expansion.
Thus a term from (\ref{eq:PSIR}) becomes:
\begin{eqnarray}
a_{nk} \left(-\frac{\zeta}{n}\right)^k
\frac{\mathrm{d}^k}{\mathrm{d}\zeta^k} \mathcal{L}_{I_\Phi}(\zeta)&
=& a_{nk}
n^{\frac{2}{\alpha}-k}\lambda\zeta^{\frac{2}{\alpha}}C_\alpha
\prod_{l=0}^{k-1}(l-2/\alpha)\nonumber\\
&&+\Theta(\kappa^2)
\end{eqnarray}
so that the outage probability is given by
\begin{eqnarray}
\mathbf{P}(SIR\leq\beta)&=&\lambda\zeta^{\frac{2}{\alpha}}C_\alpha
\sum_{n}\sum_{k} a_{nk}
n^{\frac{2}{\alpha}-k}\prod_{l=0}^{k-1}(l-2/\alpha)\nonumber\\
&&+\Theta(\kappa^2)\nonumber\\
&=&\lambda\zeta^{\frac{2}{\alpha}}\frac{C_\alpha}{K_\alpha}+\Theta(\kappa^2)\:=\:\epsilon
\label{eq:TayEnd}
\end{eqnarray}
and with $\zeta=\beta R^\alpha$ and $K_\alpha$ as in (\ref{eq:K}),
solving for $\lambda$ yields the result.

For completeness, we conclude by demonstrating how thermal noise can
be included in the analysis. To include noise, $I_\Phi$ must be
replaced by $I_\Phi+\frac{1}{\rho}N_o$ and so the transform of the
distribution $f_{I_\Phi+\frac{1}{\rho}N_o}(x)$, given by
$\mathcal{L}_{I_\Phi+\frac{1}{\rho}N_o}(\zeta)=\mathcal{L}_{I_\Phi}(\zeta)\mathcal{L}_{N_o}(\zeta/\rho)$,
replaces $\mathcal{L}_{I_\Phi}(\zeta)$ in the above derivations.
This follows from the property of Laplace transforms that the
transform of the sum of independent variables is the product of the
transforms. Now the transform of the noise is
$\mathcal{L}_{N_o}(\zeta/\rho)=e^{\zeta\cdot N_o/\rho}$.
Furthermore, since
$\frac{\mathrm{d}}{\mathrm{d}\zeta}\mathcal{L}_{N_o}(\zeta)=\frac{N_o}{\rho}\mathcal{L}_{N_o}(\zeta)$
we have
\begin{equation}
\frac{\mathrm{d}^p}{\mathrm{d}\zeta^p}\mathcal{L}_{I_\Phi+N}(\zeta)
=\mathcal{L}_{N}(\zeta)\left(\sum_{k=0}^{p}{p \choose
k}\left(\frac{N_o}{\rho}\right)^{p-k}\frac{\mathrm{d}^k}{\mathrm{d}\zeta^k}\mathcal{L}_{I_\Phi}(\zeta)\right)
\end{equation}
The expression for
$\frac{\mathrm{d}^p}{\mathrm{d}\zeta^p}\mathcal{L}_{I_\Phi+\frac{1}{\rho}N_o}(\zeta)$
now replaces
$\frac{\mathrm{d}^p}{\mathrm{d}\zeta^p}\mathcal{L}_{I_\Phi}(\zeta)$
in (\ref{eq:PSIR}). Under small outage constraints, the first order
Taylor expansion of the probability of outage can be made in a
manner analogous to equations (\ref{eq:TayBegin}) through
(\ref{eq:TayEnd}) leading to:
\begin{equation}
K_\alpha=\left[\sum_{\substack{n\in\mathcal{N}\\k\in\mathcal{K}}}\sum_{j=0}^{k}{k
\choose j}\left(\frac{N_o}{\rho}\right)^{k-j}
a_{nk}n^{\frac{2}{\alpha}-j}\prod_{l=0}^{j-1}(l-2/\alpha)\right]^{-1}.
\end{equation}
Note that this expansion is only valid when outage due to the fading
of the intended signal and thermal noise is less than $\epsilon$ in
the absence of any interference.

\section{\label{sec:pfThm1}Proof of Proposition 1}

\noindent To demonstrate the above let the interfering signals and
the desired signal be Nakagami fading with different parameters
$m_i$ and $m_o$ respectively. The CCDF of the received power is:
$F^c_{S_0}(s)=e^{-m_o s}\sum_{k=0}^{m_o-1}\frac{(m_o s)^k}{k!}$ with
$m_o=1$ being the Rayleigh case. According to Theorem 1:
\begin{equation}
\mathbf{P}(SIR\geq\beta)=\left.\sum_{k=0}^{m_o-1}\frac{(-\zeta)^k}{k!}
\frac{d^k}{d\zeta^k}\mathcal{L}_{I_\Phi}(\zeta)\right|_{\zeta=m_o\beta
R^\alpha}
\end{equation}
Note that $\zeta$ now includes the fading parameter $m_o$. To
determine the Laplace transform of the shot noise process, with
$m_i$ denoting the Nakagami parameter for all interfering
transmissions the MGF of each mark is altered to be
\begin{equation}
E\left[e^{-\zeta S_i |x|^{-\alpha}}\right]=\frac{1}{(1+\zeta/
m_i|x|^{\alpha})^{m_i}}
\end{equation}
and the integral in (\ref{eq:lapl}) can be evaluated as
\cite{IEEEbib:gr}
\begin{eqnarray}
\label{eq:lapostbc} \mathcal{L}_{I_\Phi}(\zeta)&=&\mathrm{exp}
\left\{-2\pi\lambda\int_{0}^{\infty}u\left(1-\frac{1}{(1+
u^{\alpha}/\zeta)^{m_i}}\right)\mathrm{d}u\right\}\nonumber\\
&=&\mathrm{exp}
\left\{-2\pi\lambda\int_{0}^{\infty}\frac{\sum_{k=1}^{m_i} {m_i
\choose k}u^{k\alpha+1}(m_i/\zeta)^k}{(1+
u^{\alpha}m_i/\zeta)^{m_i}}\mathrm{d}u\right\}\nonumber\\
&=&\mathrm{exp}\left\{-\lambda C_{\alpha,m_i}
\left(\frac{\zeta}{m_i}\right)^{\frac{2}{\alpha}}\right\}
\end{eqnarray}
where $C_{\alpha,m_i}=C_{\alpha,m}$ in (\ref{eq:CP}). By Theorem 1,
the optimal contention density is:
\begin{equation}
\label{eq:lamnak}
\bar{\lambda_\epsilon}=\frac{m_i^{\frac{2}{\alpha}}K_{\alpha,m_o}\epsilon}
{m_o^{\frac{2}{\alpha}}C_{\alpha,m_i}\beta^{\frac{2}{\alpha}} R^2}
\end{equation}
where $K_{\alpha,m_o}$ is given by (\ref{eq:K}). If $m_o$ is set to
$1$ (Rayleigh fading) with $K_{\alpha,1}=1$, and
$m_i\rightarrow\infty$, the MGF of the power fading mark on each
interferer approaches $e^{-\zeta|x|^{-\alpha}}$. Hence,
\begin{eqnarray}
\mathcal{L}_{I_\Phi}(\zeta)&=&\mathrm{exp}
\left\{-2\pi\lambda\int_{0}^{\infty}x(1-e^{-\zeta|x|^{-\alpha}})dx\right\}\nonumber\\
&=&\mathrm{exp}\left\{-\pi\lambda\zeta^{\frac{2}{\alpha}}\Gamma(1-2/\alpha)\right\}
\end{eqnarray}
indicating that
\begin{equation}
\lim_{m\rightarrow \infty}
\frac{C_{\alpha,m}}{m^{\frac{2}{\alpha}}}=\pi\Gamma(1-2/\alpha).
\end{equation}

If in addition $m_i=\infty$ with
$C_{\alpha,\infty}=\pi\Gamma(1-2/\alpha)$ and $m_o$ is allowed to
approach infinity, the distribution of $S_0$ becomes an impulse at
$S_0=1$. Weber, \emph{et al.} \cite{IEEEbib:tc}, derived bounds on
the optimal contention density for path loss only (non-fading):
\begin{equation*}
\left(\frac{\alpha-1}{\alpha}\right)\frac{\epsilon}{\pi
\beta^{\frac{2}{\alpha}}
R^2}\leq\bar{\lambda_\epsilon}\leq\frac{\epsilon}{\pi
\beta^{\frac{2}{\alpha}} R^2}\cdot
\end{equation*}
This gives
\begin{equation}
\left(\frac{\alpha-1}{\alpha}\right)\frac{1}{\pi}\leq\lim_{m\rightarrow
\infty}\frac{K_{\alpha,m}}{C_{\alpha,m}}\leq\frac{1}{\pi}
\end{equation}
which for fixed $C_{\alpha,\infty}$ determines the asymptotic
orderwise increase of $K_{\alpha,m}$: $\lim_{m\rightarrow \infty}
\frac{K_{\alpha,m}}{m^{\frac{2}{\alpha}}}=c_1$, for some finite,
nonzero constant $c_1$. To fully demonstrate the orderwise behavior
of $K_{\alpha,m}$, the bounds
\begin{equation}
1\leq \frac{K_{\alpha,m}}{m^{\frac{2}{\alpha}}} \leq c_1
\end{equation}
hold since $\frac{K_{\alpha,m}}{m^{\frac{2}{\alpha}}}$ is
monotonically increasing but approaches the limit $c_1$.

Equation (\ref{eq:lamnak}) is more general than (\ref{eq:naklamthm})
for which $m_o=m_i=m$, but while the physical significance of
modeling this disparity between desired and interfering statistics
is dubious\footnote[1]{For moderately dense ad hoc or sensor
networks with narrowband transmission, the key interferers are the
nearest ones, and it is unlikely that the propagation statistics of
the interfering and desired signals would be widely different.}, it
allows the behavior of $K_{\alpha,m}$ and $C_{\alpha,m}$ to be
studied. It was shown in \cite{IEEEbib:tc} that the upper bound is
fairly tight which implies that in fact $\lim_{m\rightarrow
\infty}\frac{K_{\alpha,m}}{C_{\alpha,m}}\approx\frac{1}{\pi}$. While
the upper bound holds, we have numerically that the ratio
$\frac{K_{\alpha,m}}{C_{\alpha,m}}$ does in fact approach the upper
bound with increasing $m$. Approximating closely the limit
$\frac{K_{\alpha,\infty}}{C_{\alpha,\infty}}$ as $\frac{1}{\pi}$, we
can approximate $c_1$ very closely as $c_1\approx
\Gamma(1-2/\alpha)$.

\bibliographystyle{IEEEtran}
\bibliography{Hunter}

\end{document}